# Multi-Aircraft Scheduling Optimization in Urban Environments

Jin Zhang [a], Xiaoran Qin [a, *], Ming Zhang [b]

[a] *Department of Transportation Engineering, South China University of Technology, Guangzhou, China*
[b] *College of General Aviation and Flight, Nanjing University of Aeronautics and Astronautics, Nanjing, China*

**Abstract:** With the increasing development of intelligent transportation systems and advancements in aviation technology, the concept of Advanced Air Mobility (AAM) is gaining attention. This study aims to improve operational safety and service quality within Urban Air Mobility (UAM) through a trajectory-based operation (TBO). A multi-layer operational risk assessment model is introduced to capture the effects of aircraft failure scenarios on critical urban entities, including ground personnel, vehicles, and in-flight UAVs (unmanned aerial vehicles). Based on this, a single-aircraft track planning model is designed to balance operational risk and transportation cost under the performance constraints of eVTOL (electric Vertical Take-off and Landing) aircraft. A customized track planning algorithm with safety buffer zones is used to identify the most efficient flight paths. Additionally, a multi-aircraft scheduling optimization model is proposed to minimize delays and reduce mid-air collision risks. Experimental results show that the presented approach improves both efficiency and safety, providing practical solutions for UAM operations.

## 1. Introduction

Intelligent Transportation Systems (ITS) have increasingly become a crucial component of the transportation sector, influencing roadways, railways, and air travel. At the forefront of this transformation is Urban Air Mobility (UAM), a concept in air transit proposed by organizations such as Uber, NASA, Airbus, and Ehang. UAM is redefining on-demand travel by establishing a network that integrates seamlessly with ground road traffic and underground rail transit. This innovative approach has attracted significant interest from the global air transport industry, as it offers secure, automated, and cost-effective services, particularly for urban low-altitude airspace.

Urban Air Mobility (UAM) is transforming transportation in urban areas, providing connections over short-to-medium distances. These systems operate at altitudes between 100 to 1000 meters and cover distances from 3 to 100 kilometers. eVTOL aircraft are the vehicles used in the Unmanned Aerial System (UAS) of UAM, designed to carry 2-5 passengers per trip. The advanced design of these aircraft includes distributed electric propulsion, enabling all-electric systems for intra-city travel and hybrid-electric systems for longer inter-city travel. Their ability to hover and cover long distances equips eVTOL aircraft for fully automated flights, while reducing pollution and enhancing passenger privacy. With these features, eVTOL aircraft are poised to revolutionize urban air mobility by surpassing traditional helicopters in safety, economic efficiency, comfort, and environmental impact. As they become integrated into transportation networks, their application is expected to expand to regular services such as airport transfers and interurban travel, providing tangible benefits even in less densely populated regions.

This study focuses on optimizing multi-aircraft scheduling in an urban airspace environment, integrating risk assessment, track planning, and scheduling optimization for Urban Air Mobility (UAM). An operational risk assessment model is introduced, and single-aircraft tracks are designed. Building on this, a conflict-free multi-aircraft scheduling method is proposed, which evaluates total flight numbers and average delays. Validation experiments confirm that the developed models and algorithms improve performance in optimizing urban air mobility operations within a simulated environment. The contributions of this study are summarized as follows:

- A multi-layer urban airspace environment model is developed, accounting for the impacts of aircraft on key groups, including ground personnel, vehicles, and small multi-rotor UAVs. This model effectively portrays the distribution of risk posed to them.
- The single-aircraft track planning model balances static operational risk and transportation cost, with a customized algorithm that includes a safety buffer zone to prevent safety clearance violations at upcoming waypoints.
- A multi-aircraft scheduling optimization model is developed to minimize average delay times while avoiding dynamic collisions. A tailored scheduling decision process, supported by a solution algorithm, helps alleviate airspace congestion.

The paper is organized as follows. Section 2 provides a review of existing studies, including research on urban low-altitude airspace, single-aircraft track planning, and multi-aircraft scheduling optimization. Section 3 outlines the problem statement, which forms the basis for the subsequent discussions. Section 4 elaborates on the urban airspace environment model and the models for single-aircraft track planning and multi-aircraft scheduling. Section 5 introduces the methodologies for solving the models. Section 6 focuses on verification and analysis of the results. Finally, Section 7 summarizes the conclusions and suggests directions for future work.

## 2. Related works

By reviewing the relevant literature, the existing research has been synthesized into three primary areas: urban airspace environment studies, single-aircraft track planning, and multi-aircraft scheduling.

*2.1 Urban low-altitude airspace research*

Research on urban low-altitude airspace is still in its early stages, and there is no consensus on urban airspace planning. The primary focus has been on defining the boundaries of low-altitude airspace, designing its structure, and assessing operational factors. However, these theoretical studies have yet to be validated through practical application.

In terms of defining airspace boundaries, Geister et al. [1] suggested that UAVs operate in unregulated airspace and proposed a grid system to enable aircraft with similar performance to share airspace efficiently. In contrast, NASA [2] defined low-altitude airspace as below 400ft and planned to integrate UAS Traffic Management (UTM) into the existing controlled airspace system. Regarding airspace structure design, Hoekstra et al. [3] introduced four different urban airspace structures—fully hybrid, layered, zoned, and piped—designed for UAV and PAV operations. Adding to this, the layered airspace model [4-5] was suggested as a way to balance operational safety and capacity, offering a blend of structured and unstructured airspace designs. Furthermore, Balakrishnan [6] recommended operational concepts such as basic flight, free route, corridor, and fixed route operations for UAVs.

In terms of safety management, geographical fencing [7-10] has been suggested to redirect aircraft away from high-risk areas such as city centers and airports, although this may not be feasible in cities with already limited airspace. Pang et al. [11] classified airspace into different risk categories to guide aircraft into safer zones, while Oh et al. [12] focused on risk assessment within urban environments. Besides, the FAA [13] emphasized that safety for people, vehicles, and property should be the top priority in managing low-altitude airspace. Recognizing that it is impossible to eliminate risk entirely, they put forward strategies to minimize the likelihood and severity of accidents.

In summary, previous research on urban airspace has not fully addressed operational risks or comprehensively analyzed flight safety implications. Since the success of urban air mobility depends on the precise demarcation of airspace boundaries, relying solely on subjective experience for airspace planning is insufficient to ensure safety and convenience. Thus, a key challenge remains in defining safe urban low-altitude airspace through the lens of operational risk.

*2.2 Single-aircraft track planning research*

Flight track planning is a critical component of transportation services, with significant advancements in areas such as obstacle avoidance, operational risk management, aircraft performance, and intelligent heuristic algorithms. These developments provide a solid foundation for improving flight path planning in urban low-altitude airspace.

Innovative solutions in obstacle avoidance have emerged from recent research. For example, Zhou et al. [14] and Chen et al. [15] developed path planning algorithms that enable UAVs to navigate around both static and dynamic obstacles. Tang et al. [16] proposed a low-altitude airspace management system capable of automatically generating a route network that avoids obstacles and obstructions. Building on this, Wakabayashi et al. [17] designed a collision avoidance model that considers both the position and velocity of UAVs relative to obstacles, enhancing the safety of autonomous flights in densely populated urban areas.

Operational risk is another major focus. Weibel et al. [18] formulated a ground risk assessment model based on specific events, considering factors such as aircraft system failure rate, fatal fragment coverage area, regional population density, shielding coefficient, and crash mitigation measures, thereby creating a framework for ground risk assessment. Filippis et al. [19] introduced a risk map based on ground topography and employed various algorithms for path planning. Washington et al. [20] divided the ground risk model into fault, impact, recovery, pressure, exposure, accident, and injury models, comparing their advantages and disadvantages in terms of uncertainty, assumptions, and model characteristics. In addition, Primatesta et al. [21] explored both pre-flight and real-time planning methods, focusing on risk mitigation and developing visual risk maps for urban flight.

In the realm of intelligent heuristic algorithms, Shin et al. [22] adapted these algorithms to meet UAV navigation needs in hostile environments, considering factors such as radar detection and UAV dynamics. Huang [23] improved the particle swarm optimization algorithm by incorporating a dynamic divide-and-conquer tactic, which breaks down complex high-dimensional problems into manageable low-dimensional segments. Similarly, Tong et al. [24] implemented a multi-objective track planning strategy that optimized path length, curvature, and operational risk using a hybrid algorithm that combined pigeon heuristic optimization with differential evolution. The Pareto principle was applied to determine the most efficient route, and this technique was validated through extensive testing in various scenarios. Lin et al. [25] and He et al. [26] developed trajectory planning methods for UAVs in challenging urban environments, while Andres et al. [27] used an augmented random search algorithm to iteratively deform a directed graph in the search for an optimal path.

Weather conditions also play a significant role in track planning. Hu et al. [28] advanced a weather scenario generation algorithm that maps ensemble-based weather forecast data to airspace blockage maps, and they formulated a nonlinear two-stage stochastic programming model to find optimal paths for UAV missions under weather uncertainty.

In summary, most existing algorithms primarily aim to minimize distance in continuous space, often neglecting factors such as uncertainty and convergence speed. Although some algorithms combine continuous and discrete approaches, their application in three-dimensional spaces remains limited, and they typically do not address grid-based environments. Furthermore, the constraints in these models do not sufficiently represent the unique performance characteristics of eVTOL aircraft. Therefore, defining performance constraints for eVTOL aircraft and enhancing current algorithms to identify optimal flight paths are key challenges that this paper tackles.

*2.3 Multi-aircraft scheduling optimization research*

Research on multi-aircraft scheduling optimization primarily focuses on resolving conflicts and minimizing delays to enhance safety and economic efficiency. Much of the existing research is geared toward major civil aviation airlines, with an emphasis on reducing fuel consumption to lower operational costs. Nevertheless, there is a noticeable gap in studies targeting the unique requirements of UAS (Unmanned Aerial Systems) in general aviation, which remain relatively underexplored.

In the domain of conflict resolution, Qian et al. [29] scheduled UAV takeoff times based on pre-defined tracks and speeds to proactively eliminate potential conflicts. Similarly, Wang et al. [30] evaluated the collision risk posed by intruding UAS within airport-restricted "terminal airspace," employing probabilistic conflict prediction via Monte Carlo simulations. Shihab et al. [31] proposed a hybrid model that combined on-demand with scheduled services, utilizing mixed-integer quadratic programming to manage conflicts while ensuring profitability. Building on these theoretical advancements, Zhao et al. [32] introduced a novel method for aircraft conflict resolution in air traffic management (ATM) using physics-informed deep reinforcement learning. Dong et al. [33] offered a solution grounded in the Markov decision process, employing an independent deep Q-network algorithm to handle multi-aircraft conflicts effectively. Expanding further, Chen et al. [34] recommended a general multi-agent reinforcement learning approach for real-time three-dimensional multi-aircraft conflict resolution.

Regarding delay reduction, Pradeep et al. [35] combined heuristic algorithms with mixed-integer linear programming and a time-marching algorithm capable of real-time adjustments to optimize aircraft landing schedules. Bertram et al. [36] introduced a novel approach for managing aircraft in the terminal areas of vertical take-off and landing (VTOL) airports, significantly reducing overall delays in high-density airspace by employing loops and gates to direct traffic, along with a self-organizing terminal sorting algorithm based on the Markov decision process. Furthermore, Brittain et al. [37] developed an advanced hierarchical deep reinforcement learning algorithm that sequences aircraft and ensures smooth operations by optimizing rerouting and speed adjustments during scheduling. Similarly, Lin et al. [38] proposed a multi-layer air traffic network delay model and developed a data-driven approach to capture the impact of en-route congestion on flight delays. Zhang et al. [39] introduced a local collision resolution method based on the Markov decision process for real-time dynamic collision avoidance. Pang et al. [40] put forward a framework formulated as a double-layer optimization problem, combining scheduling, speed adjustments, and rerouting strategies for conflict resolution. Lastly, Pang et al. [41] introduced a machine-learning-enhanced methodology for aircraft landing scheduling, where data-driven machine learning models were employed to improve automation and safety.

To summarize, while existing research has advanced scheduling optimization techniques, conflict resolution—especially for UAVs performing specific, non-routine tasks—remains an unresolved challenge. Furthermore, the body of work focused on multi-aircraft flight information management is still sparse, presenting obstacles to effectively reducing congestion and delays in large-scale operations. Therefore, a practical multi-aircraft scheduling optimization approach, specifically designed for the operational characteristics of urban air mobility, is critically needed.

## 3. Problem statement

To support the expansion of UAM, it is critical to tackle the diverse operational challenges related to risk management, airspace congestion, and scheduling efficiency. As UAM continues to evolve, achieving a balance between safety and efficiency becomes increasingly important, particularly in multi-aircraft scheduling. This study presents a framework that integrates both safety and efficiency. First, an operational risk model was introduced to evaluate risks posed to ground personnel, vehicles, and UAVs. Based on these evaluations, a realistic airspace model was constructed. Taking eVTOL performance into account, a track planning algorithm was developed to consider operational risks, transportation costs, and safety buffer zones, enhancing track planning efficiency. Building on this foundation, a conflict-free multi-aircraft scheduling method was proposed to optimize flight sequences and reduce delays. Experimental results demonstrated the effectiveness of the model in operational risk assessment, track planning, and scheduling optimization. The main research framework is illustrated in Figure 1.

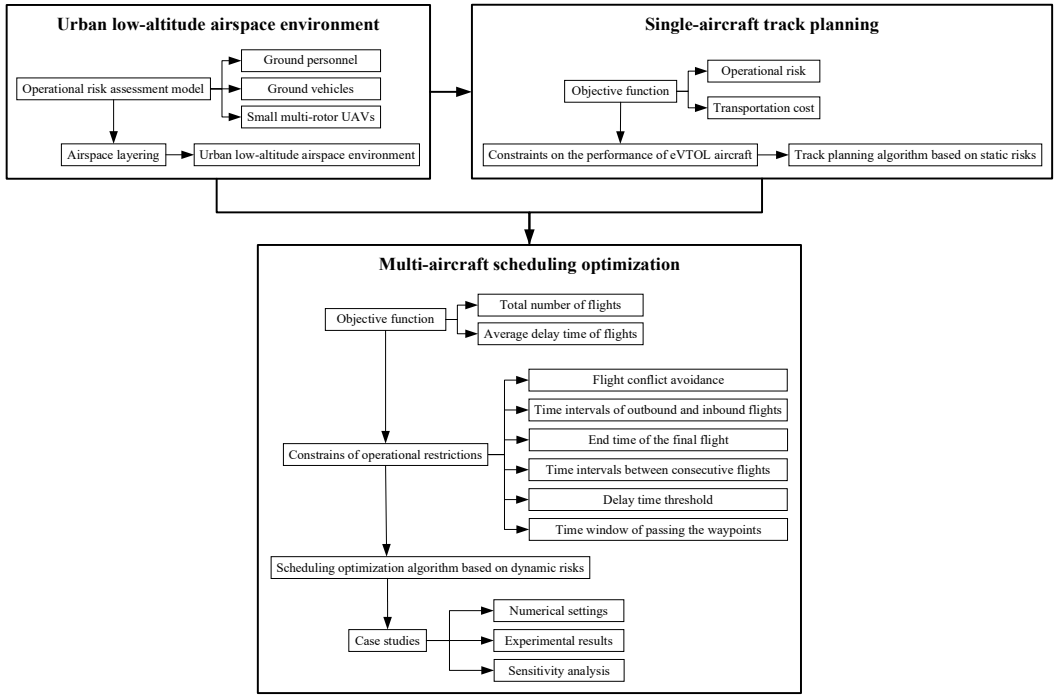

Figure 1 Framework for multi-aircraft scheduling optimization

The problem specifications and assumptions are delineated as follows:

(1) The densities of ground personnel, vehicles, and small multi-rotor UAVs are assumed to be uniformly distributed. Additionally, the speed of small multi-rotor UAVs is considered constant [42].

(2) The track planning process only considers the cruise phase between origins and destinations, excluding vertical trajectories involved in take-off and landing.

(3) The eVTOL aircraft is assumed to operate under normal environmental conditions, without interference from extreme weather or air traffic management, and without intermediate stops, maintaining a constant flight speed.

A spatial Cartesian coordinate system is employed to describe the airspace environment, with $O$ as the origin and three perpendicular axes—$x$, $y$, and $z$. The respective divisions and ranges of these axes are denoted by $\Delta x$, $\Delta y$, $\Delta z$, and $x_{max}$, $y_{max}$, $z_{max}$ respectively. The maximum indices in each direction, $a$, $b$, and $c$, are calculated by $a = \frac{x_{max}}{\Delta x}$, $b = \frac{y_{max}}{\Delta y}$, and $c = \frac{z_{max}}{\Delta z}$. The airspace is then partitioned into several identical cells, as shown in Figure 2. Each cell is located at a specific row $i$, column $j$, and layer $k$, and the center point of each cell is represented by the coordinates $(x, y, z)$. Here, $i$, $j$, and $k$ are the serial numbers of cells along the $x$, $y$, and $z$ axes, respectively. The center point coordinates of cell $(i,j,k)$ are computed as: $x = (i - 1/2)\Delta x$, $y = (j - 1/2)\Delta y$, $z = (k - 1/2)\Delta z$. For example, the center coordinates of the cell corresponding to $i = 2$, $j = 3$, and $k = 4$ are $(1.5\Delta x, 2.5\Delta y, 3.5\Delta z)$. This precise calculation of cell coordinates is essential for understanding the spatial framework in which air traffic operates and serves as the foundation for subsequent modeling and optimization efforts.

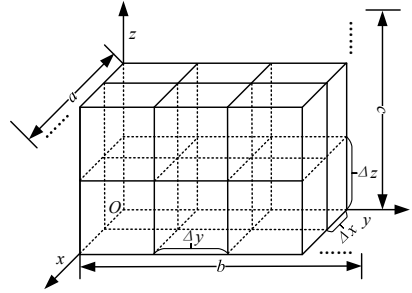

Figure 2 Schematic of airspace environment cells

## 4. Models formulation

This section presents the formulation of a multi-aircraft scheduling optimization model that prioritizes both safety and efficiency. The structure is as follows: First, the urban airspace environment model is described. Next, the single-aircraft track planning model is discussed. Finally, the multi-aircraft scheduling optimization model is introduced.

*4.1 Urban airspace environment model*

### 4.1.1 Operational risk assessment model

The operational risk assessment evaluates the specific risks posed by UAM operations to ground personnel, vehicles, and small multi-rotor UAVs. This approach, combined with airspace layering, provides a structured methodology for systematically assessing potential hazards caused by aircraft malfunctions and crashes. The risk values $R$ are determined based on the number of affected groups if an aircraft crash occurs. Three groups are considered: ground personnel, ground vehicles, and small multi-rotor UAVs. The risk to each group is calculated by considering the failure probability of the aircraft $\lambda$, the number of individuals or assets within the crash area $N$, and the probability of death or damage $P$.

The risk value for each group in a given airspace cell is expressed as:

$$R^\zeta = \lambda N^\zeta P^\zeta, \zeta = \{P, V, U\} \tag{1}$$

Where $\lambda$ is given, and the calculation methods and considerations for $N$ and $P$ differ for each group to account for their unique characteristics. The superscripts $P$, $V$, and $U$ correspond to ground personnel, ground vehicles, and small multi-rotor UAVs, respectively. The potential impact of aircraft crashes on these groups is depicted in Figure 3.

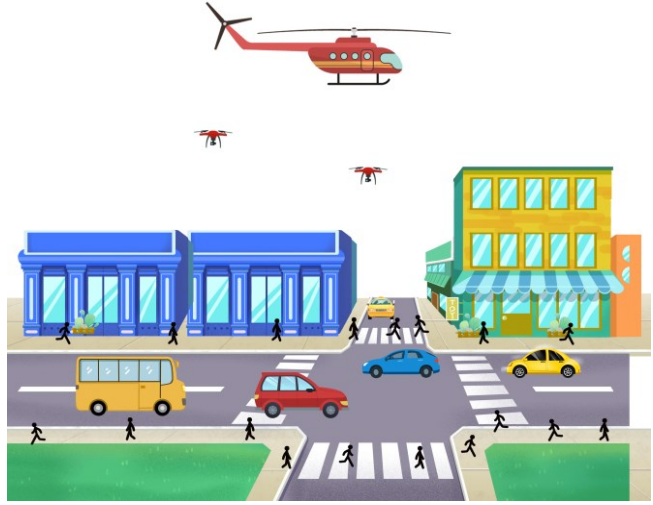

Figure 3 Schematic about the impact of aircraft falling on ground personnel, ground vehicles and UAVs

(1) Ground personnel (i.e., $\zeta = P$)

The risk to ground personnel $R^P$ from an aircraft fall is influenced by three primary factors: the probability of aircraft failure $\lambda$, the number of individuals within the crash zone $N^P$, and the probability of fatalities among those individuals $P^P$. The number of individuals at risk, $N^P$, depends on the area of the crash zone $S_e$, which is circular with a diameter $d$ equal to the length of the aircraft, and the population density within that zone $\rho_P$. The number of individuals exposed to risk in this circular area is calculated as:

$$N^P = \rho_P S_e = \frac{\rho_P \pi d^2}{4} \tag{2}$$

The probability of fatalities $P^P$ is influenced by both the sheltering factor $c_s$ and the kinetic energy of the impact $E_{imp}$. According to Dalamagkidis et al. [43], the probability of fatality can be written as:

$$P^P = \frac{1}{1+\sqrt{\frac{\alpha}{\beta}}\left(\frac{\beta}{E_{imp}}\right)^{\frac{1}{4c_s}}} \tag{3}$$

where $c_s$ is determined by the presence of trees and buildings, reducing the kinetic energy impact, and its values are given in Table B.1 of Appendix B. $\alpha$ denotes the impact energy when $c_s$ equals 0.5, and $\beta$ signifies the impact energy threshold required to cause fatality as $c_s$ approaches 0. The detailed derivation of the kinetic energy, velocity, and acceleration involved in the aircraft fall (Eq. (A.1)-Eq. (A.3)) is provided in Appendix A.

(2) Ground vehicles (i.e., $\zeta = V$)

The risk to ground vehicles $R^V$, from an aircraft fall is determined by three factors, similar to the modeling for ground personnel: the probability of aircraft failure $\lambda$, the number of vehicles within the crash zone $N^V$, and the collision probability between the aircraft and vehicles $P^V$. The number of vehicles at risk, $N^V$, is calculated as the product of the traffic density $\rho_V$ and the road length $l_r$:

$$N^V = \rho_V l_r \tag{4}$$

The probability of collision $P^V$ is derived using a two-dimensional geometric probability model. It is defined as the ratio of the total projected area of all vehicles to the total area of the road. This is described as:

$$P^V = \frac{S_v N^V}{S_r} = \frac{S_v \rho_V}{w_r} \tag{5}$$

where $S_v$ and $S_r$ signify the projected area of vehicles and road area respectively, with $w_r$ being the width of the road.

(3) Small multi-rotor UAVs (i.e., $\zeta = U$)

In addition, an out-of-control aircraft may collide with a low-flying small multi-rotor UAV involved in logistics or distribution tasks. The collision between the aircraft and the UAV is modeled in a Cartesian coordinate system, with the origin placed at the center of the aircraft. The space occupied by the aircraft is simplified into a collision box, where the length, width, and height of the box are represented by the fuselage length $e_l$, wingspan $e_w$, and aircraft height $e_h$, respectively, as displayed in Figure 4.

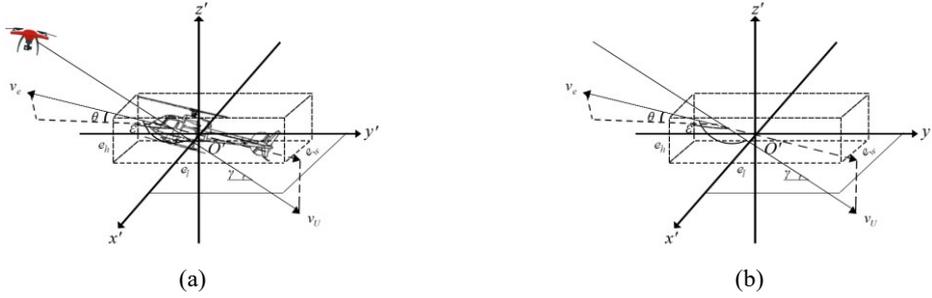

(a)            (b)

Figure 4 The impact of aircraft falling on small multi-rotor UAVs

Similar to the previous two models, the risk posed to small multi-rotor UAVs by an aircraft fall is determined by three key factors: the probability of aircraft failure leading to the fall $\lambda$, the number of UAVs within the airspace $N^U$, and the collision probability between the aircraft and UAVs $P^U$. The number of UAVs $N^U$ is calculated as the product of UAV density $\rho_U$ and the airspace volume $V$, while the collision probability $P^U$ is formulated as the ratio of the collision box volume $V_e$ to the airspace volume $V$, based on three-dimensional geometric probability models. These are denoted as:

$$N^U = \rho_U V \tag{6}$$

$$P^U = \frac{V_e}{V} \tag{7}$$

Detailed derivations of the velocity and collision probability models (Eq. (A.4)-Eq. (A.9)) are provided in Appendix A.

As a result, the operational risk value $R_{(i,j,k)}$ for cell $(i,j,k)$ indicates the cumulative risk posed by the aircraft fall to the three groups mentioned earlier. It is given by:

$$R_{(i,j,k)} = \omega_1 R^P_{(i,j,k)} + \omega_2 R^V_{(i,j,k)} + \omega_3 R^U_{(i,j,k)}, \forall i \in \{1,2,\cdots,a\}; j \in \{1,2,\cdots,b\}; k \in \{1,2,\cdots,c\} \tag{8}$$

where $\omega_1$, $\omega_2$, and $\omega_3$ are the corresponding weight coefficients for the three groups.

The safety level is typically depicted in a two-dimensional matrix, where shaded areas indicate unacceptable risk, as seen in Table B.2 of Appendix B. The value of $R_{(i,j,k)}$ can be simplified according to the following criteria:

$$R_{(i,j,k)} = \begin{cases} 1, R_{(i,j,k)} > R_0 \text{ or there are obstacles or it belongs to No-fly zone} \\ 0, R_{(i,j,k)} \leq R_0 \end{cases} \tag{9}$$

where $R_0$ is set to $10^{-7}$, defining the risk threshold for safe operation, consistent with the threshold used for manned aircraft [44].

## 4.1.2 Low-altitude airspace environment combined operational risk assessment and airspace layering

After completing the risk assessment for urban low-altitude airspace, a hierarchical zoning approach is applied within the delineated low-risk airspace. A schematic representation of the layering strategy and a detailed view of the urban low-altitude environment are presented in Figure 5 and Figure 6.

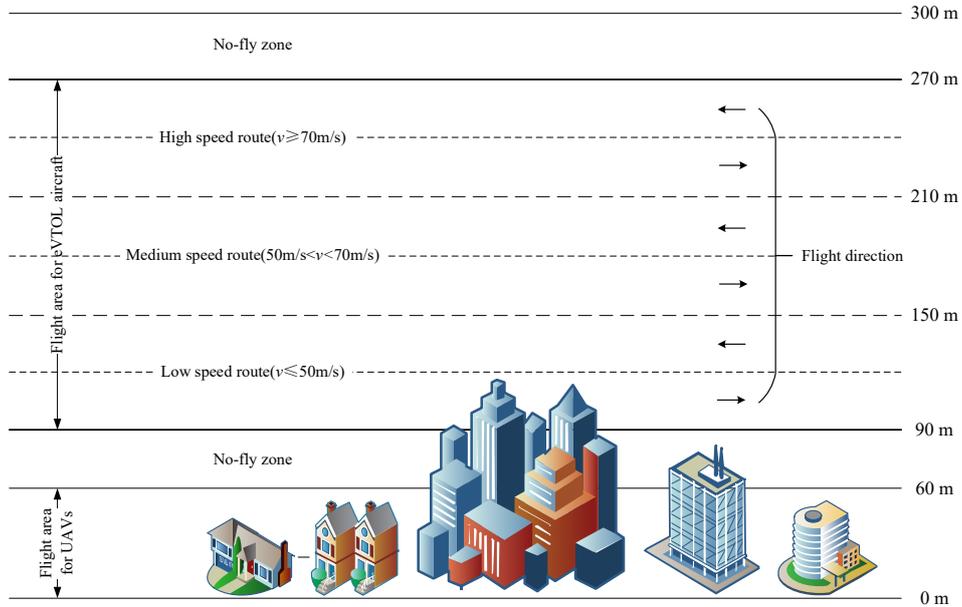

Figure 5 Schematic diagram of urban low-altitude airspace layering

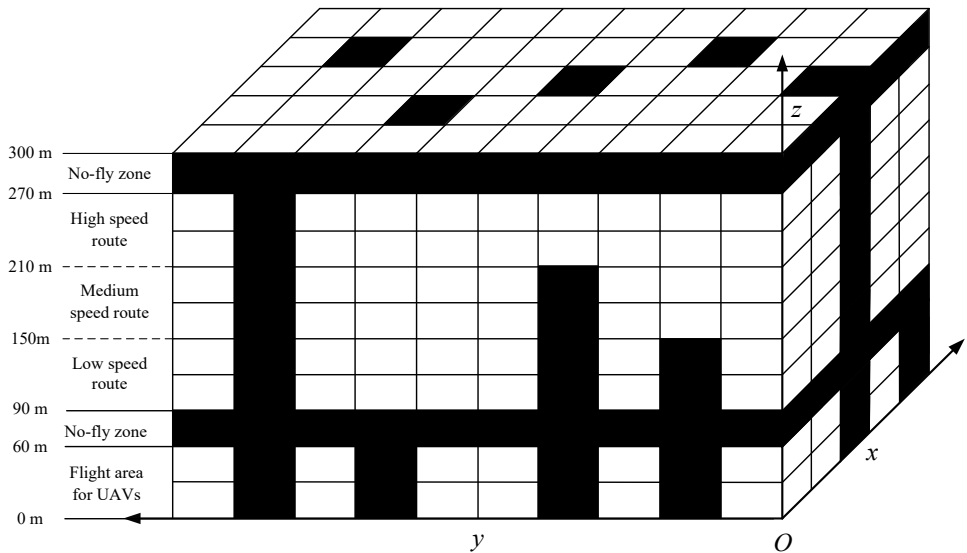

Figure 6 Schematic diagram of airspace layering and risk assessment for urban airspace environment

## 4.2 Single-aircraft track planning model

The track planning model integrates operational risk cost and transportation cost, serving as a critical prerequisite for the subsequent resolution of the multi-aircraft scheduling problem. Building upon the previously developed urban airspace environment model, the single-aircraft track planning is conducted within the constraints of the airspace and the operational capabilities of the aircraft. The decision variables in this model are the waypoints that define the track. Let the set of waypoints along the track be denoted as $P = \{p_n | n = 1, 2, \cdots, N\}$, where the spatial coordinates of the $n$-th waypoint are $p_n(i_n, j_n, k_n)$. The objective function $C$ of the single-aircraft track planning model aims to minimize a weighted sum of the operational risk cost $C_R$ and the transportation cost $C_T$, expressed as:

$$\min_{p_n} C = \omega_4 C_R + \omega_5 C_T \tag{10}$$

where $\omega_4$ and $\omega_5$ are the weighting coefficients for the operational risk cost and transportation cost, respectively. The operational risk cost $C_R$ is calculated based on the risk factors at each waypoint along the track, formulated as:

$$C_R = \sum_{n=1}^{N-1} \frac{1}{2} [R_n + R_{n+1}] \times d_{n,n+1} \tag{11}$$

where $R_n$ and $R_{n+1}$ are the risk values of the cells containing the waypoints $p_n$ and $p_{n+1}$, respectively, and $d_{n,n+1}$ denotes the Euclidean distance between them. The transportation cost $C_T$ is calculated by considering the horizontal and vertical distances of the flight segments, as well as the corresponding unit energy consumption $c_h$ and $c_v$, and unit energy cost $c_e$. It is described as:

$$C_T = \sum_{n=1}^{N-1} \left[ c_h \sqrt{(x_{n+1} - x_n)^2 + (y_{n+1} - y_n)^2} + c_v |z_{n+1} - z_n| \right] \times c_e \times \tau(m_p) \tag{12}$$

where $\tau(m_p)$ is the load factor determined by the actual passenger mass $m_p$, with the fully loaded passenger mass $m_{p\ max}$ and its corresponding load factor $\tau_{max}$ given by:

$$\tau(m_p) = 1 + \frac{m_p}{m_{p\ max}} \tau_{max} \tag{13}$$

The objective is subject to several constraints related to the performance of the eVTOL aircraft, including the flight range (Eq.(14)), waypoint requirements (Eq.(15)-(16)), and operational limitations (Eq.(17)-(19)). The flight altitude $h_n$ of the waypoint $p_n$ should remain within the range defined by both the minimum and maximum flight altitudes of the aircraft, $h_{min}$ and $h_{max}$, respectively, and the minimum and maximum flight altitudes of the designated airspace, $h'_{min}$ and $h'_{max}$. These requirements can be characterized by:

$$max\{h_{min}, h'_{min}\} \leq h_n \leq min\{h_{max}, h'_{max}\} \tag{14}$$

During the flight, the aircraft is required to maintain a safety clearance $s_{min}$ from buildings within the airspace. This is formulated as:

$$|p_n, p_b| = \sqrt{(x_n - x_b)^2 + (y_n - y_b)^2 + (z_n - z_b)^2} \geq s_{min} \tag{15}$$

where $p_b(i_b, j_b, k_b)$ refers to the center coordinates of the cells containing the buildings.

To minimize the track length, the path must avoid passing through any repeated waypoints. It can be written as:

$$p_n \neq p_{n'}, \forall p_n, p_{n'} \in P \tag{16}$$

Besides, the performance of the aircraft can influence the planning results to a certain extent. These performance-related constraints include track length, actual takeoff weight, and actual wind speed $v_{wr}$, all of which is obligated to not exceed their respective maximum values $L_e$, $m_{\max}$ and $v_{we}$. These specifications take the form of:

$$\sum_{n=1}^{N-1} \sqrt{(x_{n+1} - x_n)^2 + (y_{n+1} - y_n)^2 + (z_{n+1} - z_n)^2} \leq L_e \tag{17}$$

$$m_e + m_{p_{max}} \leq m_{max} \tag{18}$$

$$v_{wr} \leq v_{we} \tag{19}$$

*4.3 Multi-aircraft scheduling optimization model*

In air traffic management, the initial scheduling of multiple aircraft is independent yet interconnected, presenting a significant challenge in resolving potential conflicts among aircraft. To overcome this challenge, a multi-aircraft scheduling optimization model is formulated.

The objective is to minimize the weighted sum $W$ of the average delay time $T_d$ and the actual total number of flights $S$. Specifically, $T_d$ is calculated as the ratio of the total delay time of all flights to the total number of aircraft $G$, while $S$ is he total number of flights across all aircraft. This objective is achieved by determining the binary variable $c_{g,f}$ and the delay duration $d_{g,f}$ for flight $f$ from aircraft $g$. Here, $c_{g,f}$ indicates whether the flight is operational (i.e., $c_{g,f} = 1$) or cancelled (i.e., $c_{g,f} = 0$). These relationships can be articulated as:

$$\min_{c_{g,f}, d_{g,f}} W = \omega_6 T_d(c_{g,f}, d_{g,f}) - \omega_7 S(c_{g,f}) \tag{20}$$

$$T_d(c_{g,f}, d_{g,f}) = \frac{\sum_{g=1}^{G} \sum_{f=1}^{F} c_{g,f} d_{g,f}}{G} \tag{21}$$

$$S(c_{g,f}) = \sum_{g=1}^{G} \sum_{f=1}^{F} c_{g,f} \tag{22}$$

where $\omega_6$ and $\omega_7$ are the corresponding weight coefficients, while $G$ and $F$ respectively identify the number of aircraft and the number of daily planned flights. $G$ is given, and $F$ can be derived from the ratio of total service time each day and average time for a round-trip flight:

$$F = \left\lfloor \frac{T_f - T_s}{2 \times round\left(\frac{L_g}{v_g}\right) + T_D + T_R + \chi} \right\rfloor \times 2 \tag{23}$$

where $T_s$ and $T_f$ symbolize the start time and finish time of the flight schedule, respectively, $L_g$ refers to the track length, and $v_g$ is the speed of aircraft $g$. $T_D$ and $T_R$ represent the time intervals for outbound and inbound flights, respectively, and $\chi$ is the interference degree coefficient. The floor function $\lfloor \ \rfloor$ and rounding function $round()$ are used to appropriately quantize the single-flight time and round-trip times [45].

In the constraints, conflict avoidance (Eq.(24)) and time slot regulations (Eq.(25)-(31)) are considered. To ensure flight safety, which is the primary concern in multi-aircraft operations, a cell-based approach is employed to detect and prevent conflicts. Specifically, two aircraft are not allowed to occupy the same airspace cell at the same time. As illustrated in Figure 7, if two aircraft are predicted to share a cell due to overlapping tracks, intersections or shared vertiports along their routes, the slower aircraft is instructed to wait before entering the cell until the faster aircraft has departed. The mathematical formulation of this conflict avoidance strategy is as follows:

$$T_{in}(g', f', (i,j,k)) > T_{out}(g, f, (i,j,k)), v_g \geq v_{g'} \tag{24}$$

where $T_{in}$ and $T_{out}$ are the entry and exit time of the aircraft, and $v_g$ and $v_{g'}$ are the speeds of aircraft $g$ and $g'$, respectively.

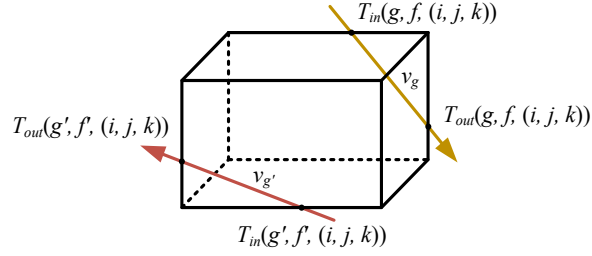

Figure 7 Schematic diagram of flight conflict between two aircraft

In addition, a series of regulations regarding the flight schedule needs to be observed. Since more resources are typically available for maintenance or charging at the origin, the time interval for outbound flights $T_D$ ought to be less than that for inbound flights $T_R$. Equally important, all aircraft is required to complete their flights and return to their origins before $T_f$. These take the following form:

$$T_D < T_R \tag{25}$$

$$A_{g,F'} + d_{g,F'} \leq T_f \tag{26}$$

where $A_{g,F'}$ and $d_{g,F'}$ are the arrival time and the corresponding delay time of the last flight $F'$ from aircraft $g$, respectively. Simultaneously, there would be a certain time interval $\Delta T$ between two consecutive flights of the same aircraft, which can be written as:

$$(D_{g,f+1} + d_{g,f+1}) - (A_{g,f} + d_{g,f}) \geq \Delta T \tag{27}$$

$$\Delta T = \begin{cases} T_D, \mod(f,2) = 1 \\ T_R, \mod(f,2) = 0 \end{cases} \tag{28}$$

where $D_{g,f}$ is the departure time of flight $f$ from aircraft $g$.

To ensure a certain degree of on-time performance in actual flight operations, a delay threshold $\Delta d_{g,f}$ is set, which cannot exceed the sum of the individual flight time and the corresponding time interval between successive flights. To align the model with actual operational requirements, the time $T_n$ for each waypoint is calculated using interpolation methods, as these times must be controlled within a specified range:

$$\Delta d_{g,f} \leq round\left(\frac{L_g}{v_g}\right) + \Delta T \tag{29}$$

$$T_n - \Delta d_{g,f} \leq T_n \leq T_n + \Delta d_{g,f} \tag{30}$$

$$T_n = \sum_{n=1}^{N-1} \frac{|p_n - p_{n+1}|}{v_g} = \sum_{n=1}^{N-1} \frac{\sqrt{(x_n - x_{n+1})^2 + (y_n - y_{n+1})^2 + (z_n - z_{n+1})^2}}{v_g} \tag{31}$$

## 5. Solution algorithm

In addressing the multi-aircraft scheduling problem, an initial flight schedule must be generated before conflicts can be resolved. The single-aircraft track planning approach assigns a feasible and safe track for each aircraft, serving as the initial input for the multi-aircraft scheduling problem. However, due to timing constraints, conflicts between multiple aircraft may arise within this initial schedule. To resolve such conflicts, a preliminary strategy is introduced, which simplifies the task by breaking down intricate, multi-aircraft conflicts into smaller pairwise conflicts, which are easier to handle and resolve.

This strategy significantly mitigates the overall problem difficulty. It also limits the number of conflicts that the solution algorithm must process, thereby narrowing the search space and lightening the computational burden. Subsequently, the scheduling optimization algorithm further refines the timing of all aircraft in the schedule. Through operations such as selection, crossover, and mutation, the algorithm iteratively improves the scheduling plan, searching for optimal or near-optimal flight sequences that minimize delays and cancellations while maintaining overall system coordination and safety.

*5.1 Track planning algorithm with safety buffer zone*

The A* algorithm is selected as the track planning strategy due to its simplicity in implementation and its guarantee of optimality. However, while it ensures the identification of the shortest path on graphs, it does not guarantee the discovery of the optimal track in a real continuous environment [46]. The standard A* algorithm may yield the best solution under specific conditions, but its computational complexity is relatively high. To tackle this, a track planning algorithm with a safety buffer zone, based on the A* algorithm, is put forward. This algorithm assumes a heuristic factor of zero, and its objective function $\phi(t)$ is defined as follows:

$$\phi(t) = \varphi(t) + \psi(t) = \eta(t, O) + \eta(t, D) \tag{32}$$

$$\eta(t, \iota) = \omega_4 C_R^\iota(t) + \omega_5 C_T^\iota(t), \iota \in \{O, D\} \tag{33}$$

Here, $\varphi(t)$ represents the actual cost function from the origin to the current waypoint $p_t$, while $\psi(t)$ signifies the estimated cost function from the current waypoint $p_t$ to the destination. Specifically, $\eta(t, O)$ and $\eta(t, D)$ symbolize the total cost from the origin to the current waypoint $p_t$, and from the current waypoint $p_t$ to the destination, respectively. Additionally, $C_R^\iota(t)$ and $C_T^\iota(t)$ correspond to the operational risk cost and the transportation cost between the waypoints $p_t$ and $\iota$, respectively.

To mitigate the risk of proximity to structures during flight, a safety buffer zone is defined. It maintains a safety margin between the aircraft and obstructions, as illustrated by the shaded area in Figure 8. The safety buffer zone consists of $n_s \times n_s \times n_s$ airspace cells, where $n_s$ can be calculated by $n_s = \left\lceil \frac{s_{min}}{s} \right\rceil \times 2 + 1$, with $s = \{\Delta x, \Delta y, \Delta z\}$, and $\lceil \ \rceil$ denoting the ceiling function. remains protected within the safety buffer zone, a penalty term $R(t)$ is integrated into the objective function if any cell on the outermost edge of the buffer zone exceeds the acceptable risk level. This prevents the aircraft from expanding to waypoints that do not satisfy the required safety clearance:

$$\phi'(t) = \varphi(t) + \psi(t) + \mu \times R(t) \tag{34}$$

$$R(t) = \sum_{k=1,n_s} \sum_{i=1}^{n_s} \sum_{j=1}^{n_s} R_{(i,j,k)} + \sum_{k=2}^{n_s-1} \left( \sum_{i=1,n_s,} \sum_{j=1}^{n_s} R_{(i,j,k)} + \sum_{j=1,n_s} \sum_{i=2}^{n_s-1} R_{(i,j,k)} \right) \tag{35}$$

where $\mu$ is the penalty coefficient. If the cell is within the safe range, $R_{(i,j,k)}=0$.

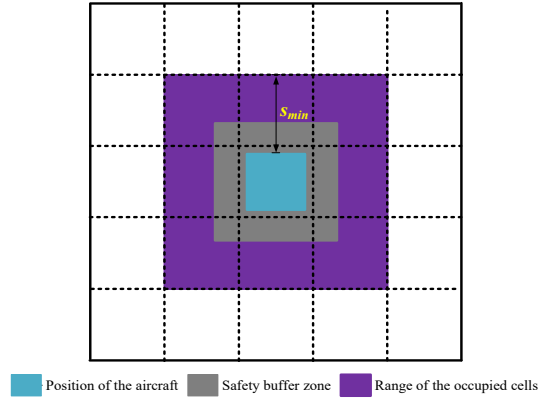

Figure 8 Planar schematic diagram of the safety buffer zone

Urban low-altitude airspace is discretized into uniformly precise cells, but the shortest path does not necessarily pass through the centers of these cells. The initial track obtained through the modified objective function defines the flight boundaries, and the corresponding equivalent track within the integrated cells allows for free navigation within these confines, as visualized in Figure 9. When bypassing the cell center is allowed, the track optimization simplifies to identifying the shortest path between adjacent integrated cells. The goal of accurate planning is therefore to determine the entry and exit nodes for each cell. The geometry of the intersection between the previous and following integrated cells differs, as do the corresponding methods for node position calculation. If the intersection is a point or a line segment, the aircraft's position upon entering the next integrated cell is $(\frac{x_{ce}+x'_{ce}}{2}, \frac{y_{ce}+y'_{ce}}{2}, \frac{z_{ce}+z'_{ce}}{2})$ or $(x_e, \frac{y_{ce}+y'_{ce}}{2}, \frac{z_{ce}+z'_{ce}}{2})$, where $p_{ce}(x_{ce}, y_{ce}, z_{ce})$ and $p'_{ce}(x'_{ce}, y'_{ce}, z'_{ce})$ are the central coordinates of the previous and following integrated cells, respectively. Similarly, $p_e(x_e, y_e, z_e)$ and $p'_e(x'_e, y'_e, z'_e)$ denote the positions where the aircraft enters the previous and following integrated cells, respectively. Here, $e$ signifies the number of integrated cells.

Nevertheless, the equivalent track derived using the revised objective function and integrated approach often results in large angles and excessive turning. To resolve this, cubic spline interpolation is employed to refine the track, enhancing its practicality. Each segment is represented by a cubic polynomial, ensuring continuity and differentiability up to the second order. The corresponding function $\vartheta(x)$ is derived by imposing boundary conditions. The optimal interpolation function $sin(x)$ is approximated using eight coordinate points surrounding the entry and exit nodes of each integrated cell, as seen in Figure 10.

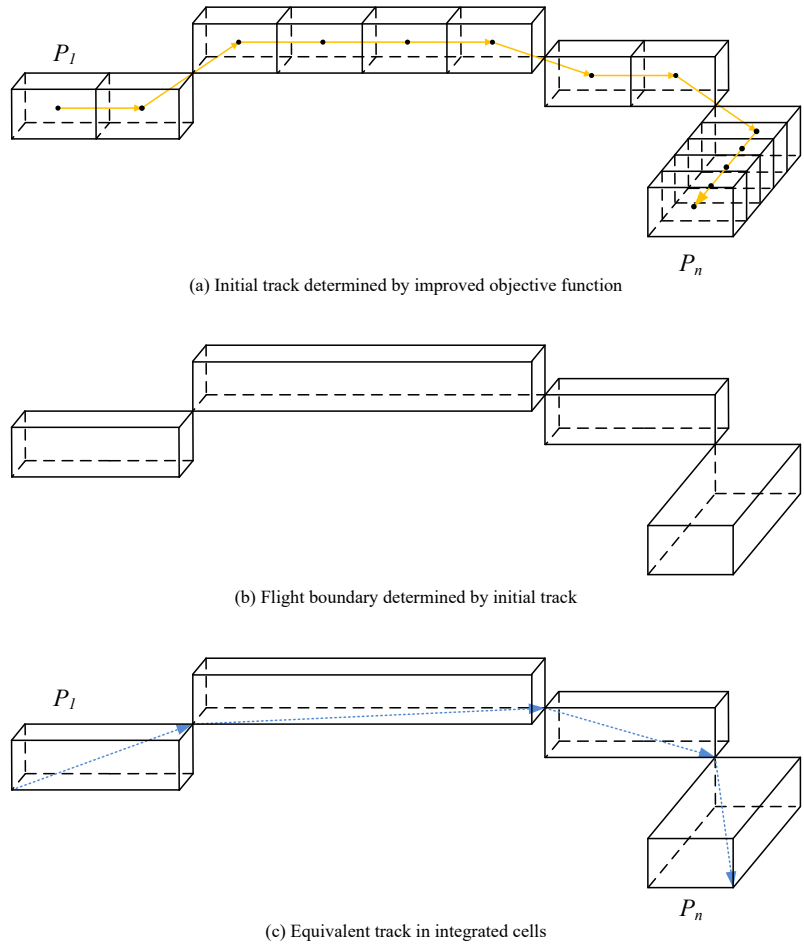

(a) Initial track determined by improved objective function

(b) Flight boundary determined by initial track

(c) Equivalent track in integrated cells

Figure 9 The process of track optimization

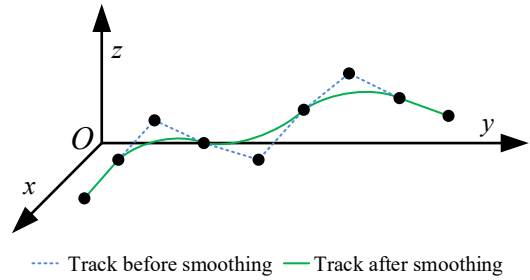

---- Track before smoothing  —— Track after smoothing

Figure 10 Schematic diagram of the track smoothing process

*5.2 Scheduling optimization algorithm based on dynamic risks*

**5.2.1 Preliminary strategy**

The primary objective is to identify the optimal scheduling solution that minimizes delays and cancellations for all flights, which necessitates resolving conflicts between multiple aircraft. However, due to the large number of flights and the high complexity involved in resolving these conflicts, it is impractical to solve the problem directly. To handle this challenge, the problem is broken down into smaller, more manageable pairwise conflicts. Simultaneously, the periodicity of urban air routes and the equal prioritization of aircraft allow for flexible arrangement of flight sequences.

The proposed method simplifies the conflict resolution process by focusing on the interactions between pairs of flights, as shown in Figure 11. An array is used to record the sequence in which flight conflicts are resolved. Here, $f'$ symbolizes the current position of flight $f$ in the conflict resolution order. Systematic checks begin with the second element in the flight list. If a conflict with any preceding flight (from $1$ to $F'-1$) is detected, the current flight $f$ is delayed. After all flights have undergone conflict checks and adjustments relative to their predecessors in the sequence, the final order is determined. The departure times for the flights are then set according to the adjusted sequence, and the multi-aircraft scheduling optimization model computes the objective function value to select the optimal sequence from all possible options.

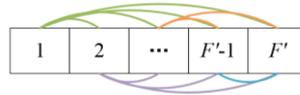

Figure 11 Schematic of the conflict resolution order in the flight sequence

In alignment with the prior methodology, the strategies for handling conflicts among flights are delineated in Figure 12:

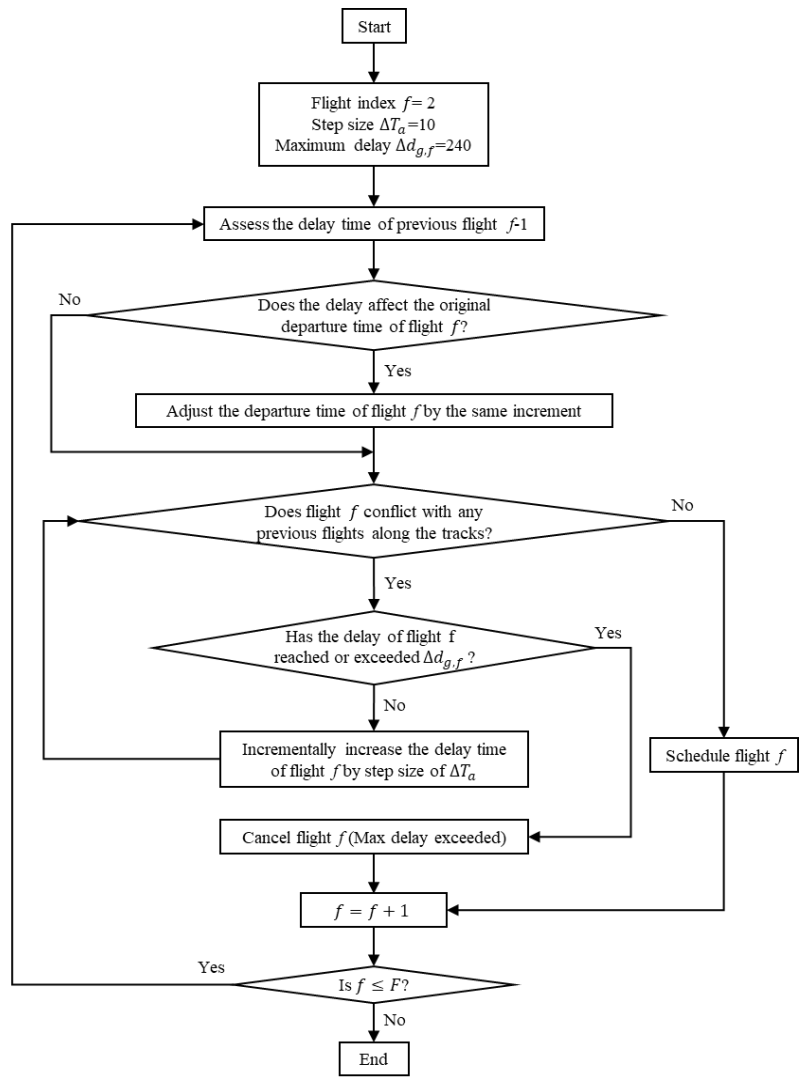

Figure 12 Flowchart of the scheduling decision process

Apart from these considerations, as discussed in Section 5.1, track optimization within an integrated cell may result in aircraft originally positioned at the cell center converging on the same surface. Alternatively, the smoothed track could briefly exit the safe region at the original turn angles, potentially leading to multi-aircraft conflicts. To tackle these issues, the number of cells in the conflict detection zone is set to $3 \times 3 \times 3$, ensuring that the distance between two aircraft remains greater than or equal to the divisions in the corresponding directions, as evident in Figure 13.

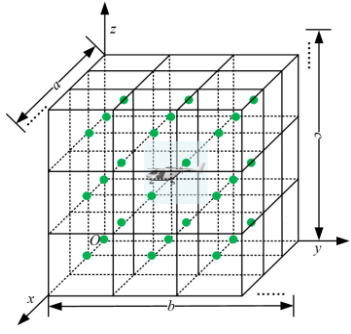

Figure 13 Order of the resolution of flight conflict

**5.2.2 GA-based improvement**

The standard genetic algorithm (GA) encounters challenges like the loss of elite individuals, slow convergence, and vulnerability to local optima. Consequently, a GA-based dynamic risks scheduling optimization algorithm (SOA) that replaces the proportion of roulette choices with the probability of a simulated annealing algorithm is devised. Furthermore, new methods for crossover and mutation are introduced to further enhance the optimization.

(1) Basic settings

In the optimization process, the chromosome utilizes an integer encoding scheme, where the integer sequence numbers correspond to the adjusted sequence numbers of the aircraft after optimization, and the integer values signify the delay times of the aircraft. Since each aircraft follows a pre-determined track with a known length, the transit times at each waypoint during the optimization process can be calculated based on the speed specified in the encoding scheme. For example, as visualized in Figure 14, aircraft $e_1$ has an actual sequence number of 1 and a delay time of 40 seconds, aircraft $e_2$ has a sequence number of $M_1 + 2$ and a delay time of 50 seconds, and aircraft $e_m$ is assigned a sequence number of $M_{m-1} + M_m$ with a delay time of 60 seconds.

| | $e_1$ | | | | $e_2$ | | | …… | | $e_m$ | | |
|---|---|---|---|---|---|---|---|---|---|---|---|---|
| Serial number | 1 | 2 | … | $M_1$ | $M_1$+1 | $M_1$+2 | … | $M_1$+$M_2$ | … | $M_{m-1}$+1 | $M_{m-1}$+2 | … | $M_{m-1}$+$M_m$ |
| Value | 40 | 0 | … | 0 | 0 | 50 | … | 0 | … | 0 | 0 | … | 60 |

Figure 14 Chromosome coding mode

To facilitate the computation of the objective function values, these values are first normalized within the range [0,1]. This normalization process is performed to derive the corresponding fitness function values, which are employed in the optimization algorithm.

$$W'^l_q = \frac{W^l_{max} - W^l_q}{W^l_{max} - W^l_{min}} \tag{36}$$

Here, $W^l_q$ stands for the objective function value of individual $q$ in generation $l$, and $W^l_{max}$ and $W^l_{min}$ correspond to the maximum and minimum values of the objective function in generation $l$, respectively. A value of $W'^l_q$ closer to 1 indicates proximity to the optimal sequence. Consequently, the fitness function can be expressed as:

$$Q_q^l = \begin{cases} W'^l_q, & \text{the individual meets the requirements} \\ 0, & \text{the individual does not meet the requirements} \end{cases} \tag{37}$$

(2) Iterative process

The iterative process involves selection, crossover, and mutation. The selection method commonly used is the simple and easy-to-implement roulette wheel selection. Even so, this method may unintentionally lead to the omission of high-quality individuals. Therefore, the concept of the simulated annealing algorithm is adopted to enhance the selection mechanism. Additionally, an elitism preservation mechanism is applied to retain individuals with higher fitness from the parent generation, thereby increasing the likelihood of selecting superior individuals. The detailed steps are outlined in Table 1.

Table 1 Pseudocode for modifying selection process by using simulated annealing concept

| | Begin |
|---|---|
| 1: | Initialize population $I(l)$ |
| 2: | Calculate the fitness values of all individuals in the population |
| 3: | If $l < l_0$ |
| 4: |    Eliminate individuals failing to satisfy the constraints |
| 5: |    Select the 10% of the population exhibiting the lowest fitness values |
| 6: |       Randomly select individuals $I_1^l$ and $I_2^l$ from the selected group $I'(l)$ |
| 7: |       Calculate their fitness value $Q_1^l$ and $Q_2^l$ |
| 8: |          If $Q_1^l > Q_2^l$ |
| 9: |             Update $I(l+1)$ with $I_2$ |
| 10: |          Else |
| 11: |             Calculate probability $P_l = e^{\frac{Q_1^l - Q_2^l}{T}}$ |
| 12: |             Update $I(l+1)$ with $I_2$ by a certain probability $P_l$ |
| 13: |       Repeat from *Step* 7 until all combinations of two individuals has been traversed |
| 14: |       End |
| 15: |    Update temperature $T = T_0 \times \xi^l$ |
| 16: | Replenish the population by selecting an equal number of individuals with the highest fitness values in $I(l)$ |
| 17: | End |

The crossover and mutation processes in the standard genetic algorithm often result in slow convergence and the algorithm getting trapped in local optima. To mitigate these limitations, both components were refined.

Both single-point and two-point crossover operators are adopted to generate offspring. In single-point crossover, a random position along the length of two parent chromosomes is selected, and all the genes from that point onwards are swapped between the two chromosomes, producing two new offspring. In two-point crossover, two random positions along the parent chromosomes are selected, and the segment of genes between these two points is swapped, also generating two new offspring. These processes are displayed in Figure 15.

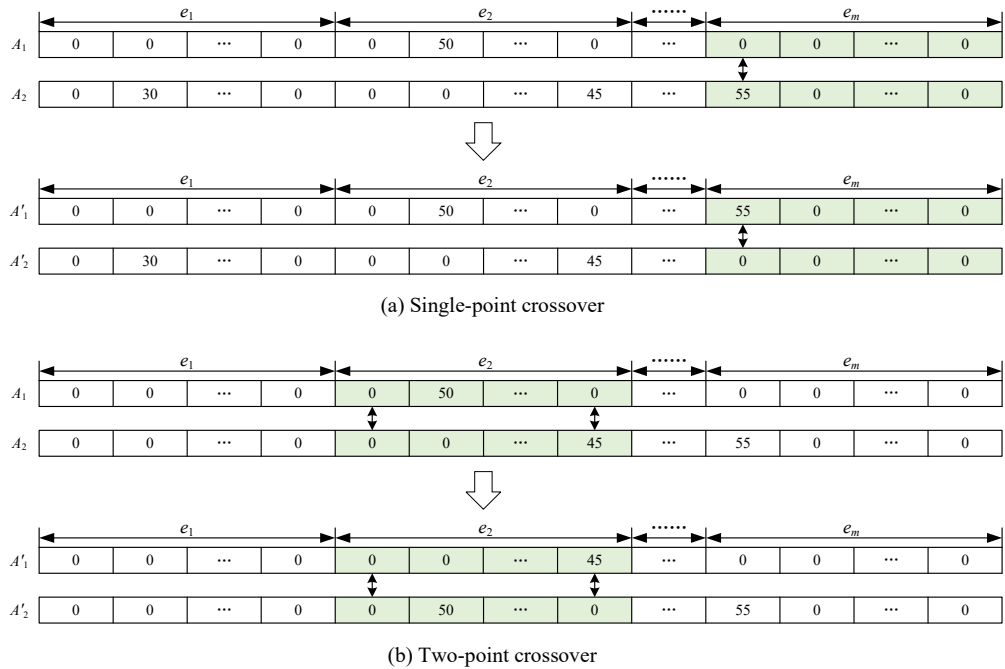

(a) Single-point crossover

(b) Two-point crossover

Figure 15 Schematic of single-point crossover and two-point crossover

The single-point crossover method was applied to individuals with fitness values below the average to maximize changes in the chromosome structure. This increased the probability of producing individuals with higher fitness. Conversely, for individuals with fitness values above the average, the two-point crossover method was utilized to preserve the original chromosome structure as much as possible, thereby protecting the superior individuals.

Both single-point and multi-point mutation strategies were integrated to increase the diversity of the population. Single-point mutation refers to the replacement of a gene in the chromosome with one of its alleles at a certain probability, where only one gene is altered. Multi-point mutation involves replacing multiple genes in the chromosome with their respective alleles at a certain probability. These processes are depicted in Figure 16.

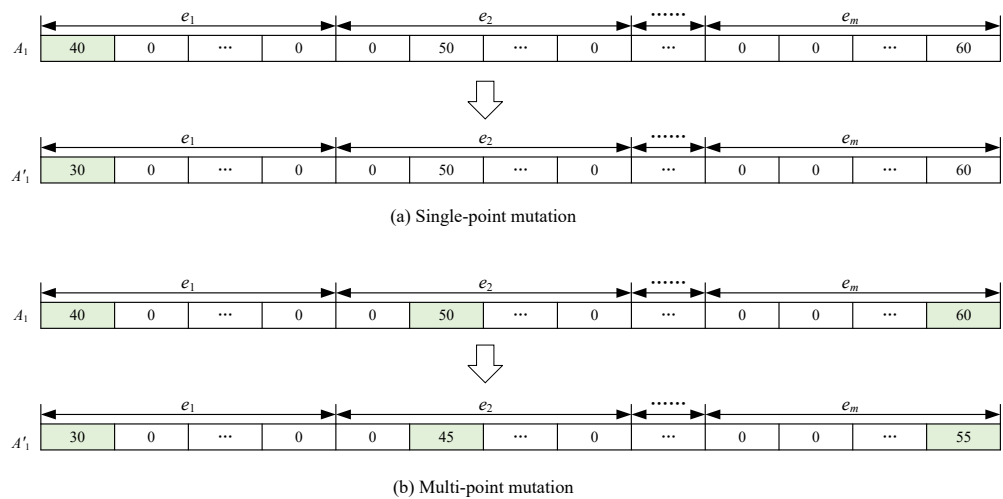

(a) Single-point mutation

(b) Multi-point mutation

Figure 16 Schematic of single-point mutation and two-point mutation

For individuals with fitness values below the average, the multi-point mutation approach is employed to maximize changes in the chromosome structure and potentially improve fitness as much as possible. In contrast, for individuals with fitness values above the average, the single-point mutation method is utilized to protect the better individuals.

The combination of refined crossover and mutation operations helps maintain reasonable crossover and mutation probabilities, as well as population diversity. This mitigates the issue of premature convergence to local optima. By applying these improved crossover and mutation procedures, the scheduling optimization algorithm based on dynamic risks enhances the robustness and efficacy of the multi-aircraft scheduling optimization process. This ensures a thorough exploration of the solution space, avoids premature convergence, and simultaneously accelerates the convergence speed of the algorithm.

## 6. Case studies

*6.1 Numerical setups*

To evaluate the effectiveness of the proposed model and algorithm, a series of simulation experiments were conducted, with the key parameters summarized in Table B.3 of Appendix B. The dataset includes several components: Digital Surface Model (DSM) data of Nanjing, China, provided by the Zhejiang Jiande General Aviation Research Institute, and performance parameters of the EH-216S eVTOL aircraft, sourced from publicly available data [47]. Moreover, simulated data of an initial flight plan with potential multi-aircraft conflicts were generated to further support the scheduling optimization.

*6.2 Experimental results*

The division accuracy of the *x*-axis, *y*-axis, and *z*-axis was set to 50m, 50m, and 30m, respectively. Using these cell dimensions, the risk for each cell in the airspace environment was computed by utilizing the previously defined operational risk assessment model. To visually convey the risk levels, a continuous color spectrum was applied, with a gradation from cooler to warmer hues representing increasing risk values. The resulting urban low-altitude airspace environment based on this operational risk assessment can be seen in Figure 17.

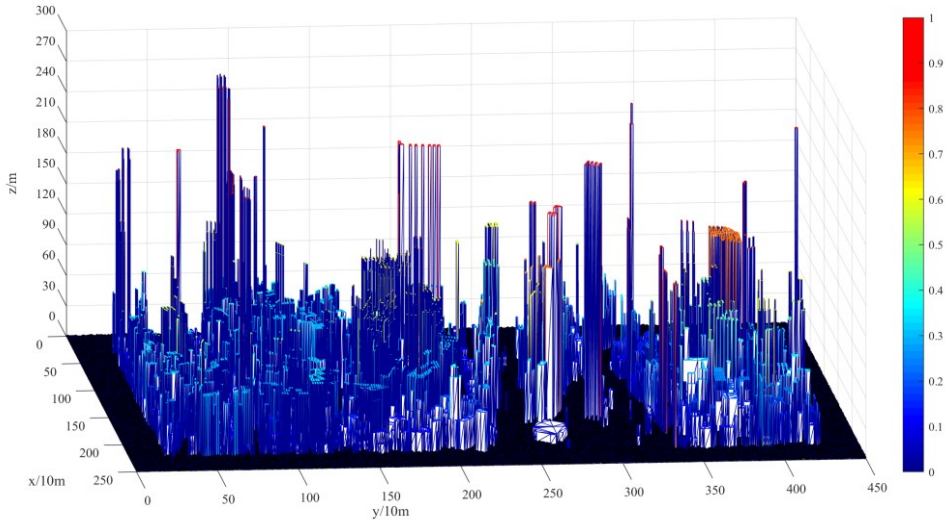

Figure 17 Urban low-altitude airspace environment after risk assessment

As detailed in Figure 18, the shortest track (red) was derived from the standard A* algorithm, which focuses on minimizing transportation costs but results in the highest risk. The initial track (orange) ensured that the aircraft passed through safer areas, but the cell-based search could lead to unnecessary traversal to the center of each cell and excessive changes in direction during actual flight. The equivalent track (yellow) refined the initial track by merging adjacent cells, reducing the number of waypoints, and straightening the track. Finally, a smoothing process was applied to reduce jagged turns, yielding a more continuous and feasible optimal track (green).

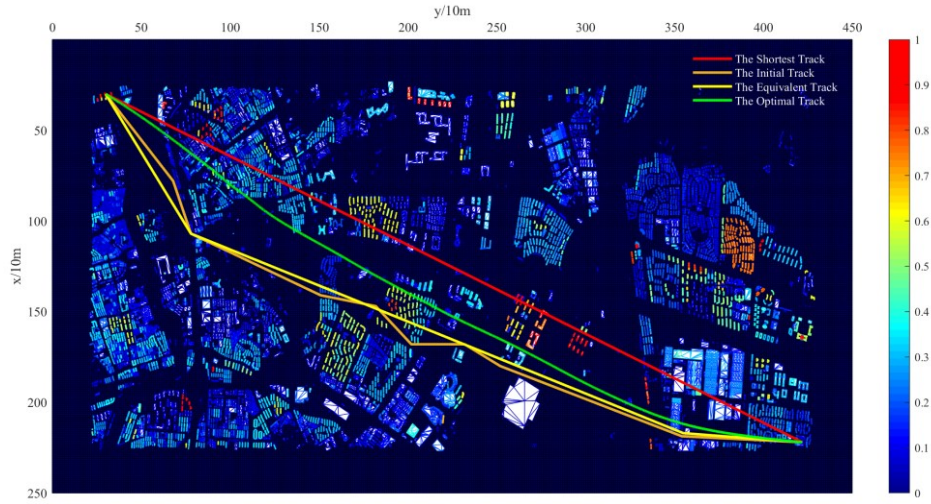

Figure 18 Tracks from different optimization procedures

Compared to the shortest track, the initial, equivalent, and optimal tracks demonstrated reductions in operational risk. However, these reductions came at the cost of increased transportation expenses and longer computational time. Detailed comparative results for these tracks are presented in Table 2. Thus, the optimal track, generated by the proposed track planning algorithm with an integrated safety buffer, significantly lowers operational risk while incurring only modest increases in cost, waypoints, and computational time compared to the shortest track produced by the A* algorithm.

Table 2 Results of track planning from different optimization procedures

| **Index** | The shortest track | The initial track | The equivalent track | The optimal track |
| --- | --- | --- | --- | --- |
| Operational risk | 196 | 174 (-11.36%) | 174 (-11.36%) | 181 (-7.69%) |
| Transportation cost(¥) | 27.1527 | 32.7379 (20.56%) | 32.0715 (18.16%) | 30.8265 (13.59%) |
| Number of waypoints | 386 | 456 (18.07%) | 453 (17.25%) | 432 (11.73%) |
| Computational time (s) | 68.71 | 90.58 (31.83%) | 101.96 (48.39%) | 108.97 (58.60%) |

Notably, the optimal track shares similarities with both the shortest track and the lowest-risk track because it strikes a balance between operational risk and transportation costs. Since transportation cost is directly related to flight distance, as the cost decreases, the optimal track naturally converges toward the shortest track. Simultaneously, the optimal track retains characteristics of the lowest-risk track due to its consideration of risk factors, ensuring that safety is not compromised for efficiency. This equilibrium between reducing transportation costs and mitigating operational risk makes the optimal track an effective compromise between the shortest and safest tracks.

Utilizing the methods described previously, a network of air routes was designed by planning 10 tracks that connected 5 vertiports within the designated airspace. This network served as the foundation for optimizing the scheduling of multi-aircraft operations. To demonstrate the practicality of these strategies, 120 flights were arranged to be executed by 20 eVTOL aircraft within this network, with an equal number of 10 flights assigned to each air route. The coordinates of each vertiport, along with detailed track information, are fully listed in Appendix A.

Each flight was assigned a specific departure time and a delay threshold. The initial plan consisted of 120 flights ($f = 120$), of which 109 actually took off, 11 were cancelled and 20 were delayed. The total accumulated delay was 2401 seconds, averaging 120.05 seconds per flight. These data reveal a certain gap between the plan and the actual operation. To mitigate this gap, a scheduling optimization algorithm based on dynamic risks was invoked. This algorithm increased the number of actual flights to 117, and reduced cancellations to 3. Notably, the total delay time of the original 20 delayed flights was reduced to 1911 seconds, with an average delay of 95.55 seconds per flight. These changes indicate a 6.67% increase in the number of actual flights and a 20.41% reduction in average delay time.

In addition to the optimization algorithm based on dynamic risks, GA was also applied. This approach increased the number of actual departures to 114, and reduced cancellations to 6. The total delay time decreased to 2161 seconds, with the average delay time per flight dropping to 108.05 seconds. This shows a 4.17% increase in the number of flights and a 10.00% reduction in the average delay time. When comparing the scheduling optimization algorithm with GA, the former outperformed the latter, with a 2.50% higher increase in actual flights and an additional 10.41% reduction in average delay time, as indicated in Figure 19. The detailed pre- and post-optimization data, including aircraft numbers, flight numbers, and relevant delay times, are recorded in Table 3. These statistics demonstrate the improved punctuality achieved through optimization and highlight the performance differences between the two algorithms in multi-aircraft scheduling.

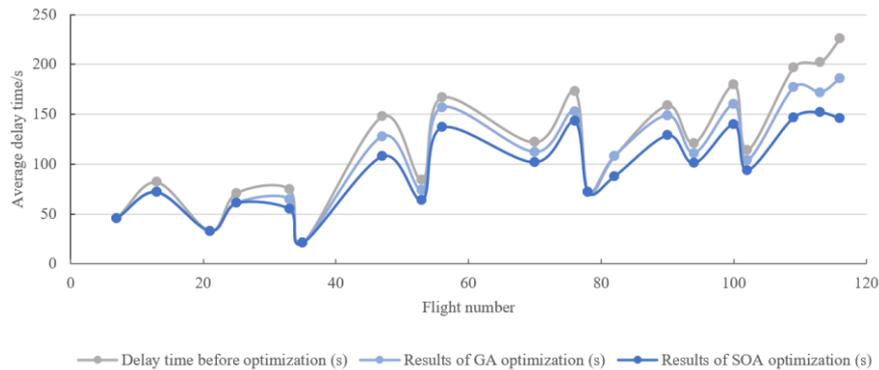

Figure 19 Comparison of delay time before and after the optimization by different algorithms

Table 3 Comparison of $d_{g,f}$ before and after optimization using different algorithms

| Flight number $f'$ | $d_{g,f}$ before optimization(s) | Results of SOA(s) | Results of GA(s) | Rate of change |
|---|---|---|---|---|
| 7 | 46 | 46 | 46 | 0% |
| 13 | 82 | 72 | 72 | 0% |
| 21 | 33 | 33 | 33 | 0% |
| 25 | 71 | 61 | 61 | 0% |
| 33 | 75 | 55 | 65 | -18.18% |
| 35 | 21 | 21 | 21 | 0% |
| 47 | 148 | 108 | 128 | -18.52% |
| 53 | 84 | 64 | 74 | -15.63% |
| 56 | 167 | 137 | 157 | -14.60% |
| 70 | 122 | 102 | 112 | -9.80% |
| 76 | 173 | 143 | 153 | -6.99% |
| 78 | 72 | 72 | 72 | 0% |
| 82 | 108 | 88 | 108 | -22.73% |
| 90 | 159 | 129 | 149 | -15.50% |
| 94 | 121 | 101 | 111 | -9.90% |
| 100 | 180 | 140 | 160 | -14.29% |
| 102 | 114 | 94 | 104 | -10.64% |
| 109 | 197 | 147 | 177 | -20.41% |
| 113 | 202 | 152 | 172 | -13.16% |
| 116 | 226 | 146 | 186 | -27.40% |

*6.3 Sensitivity analysis*

**6.3.1 Impact of flight altitude on the safety of the urban airspace environment**

The urban landscape was held constant while examining the effects of varying flight altitudes on risk assessment. The risk assessment results for the urban airspace environment at altitudes of 100m, 160m, and 220m for high-speed routes are depicted in Figure 20. The results indicate that a moderate increase in altitude leads to a reduction in risk. Specifically, the dispersion of safe airspace at lower altitudes increases risk, while the consolidation of accessible airspace at higher altitudes reduces risk, thereby improving the availability of airspace for safer and more efficient operations.

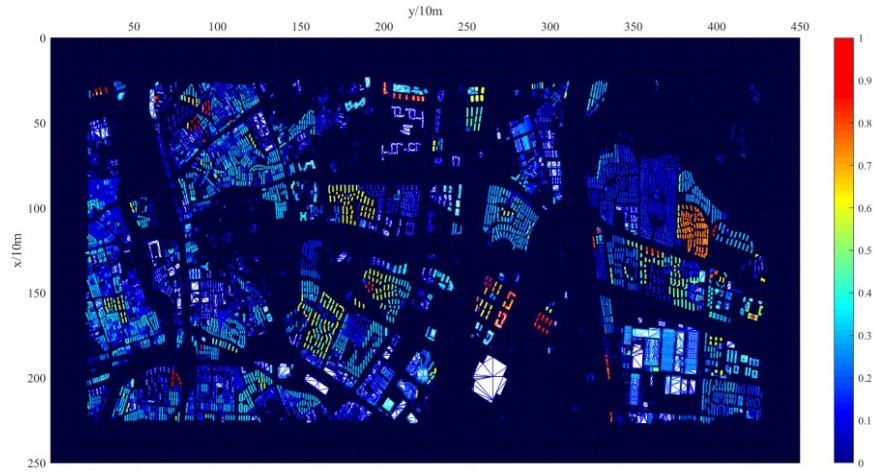

(a) h=100m

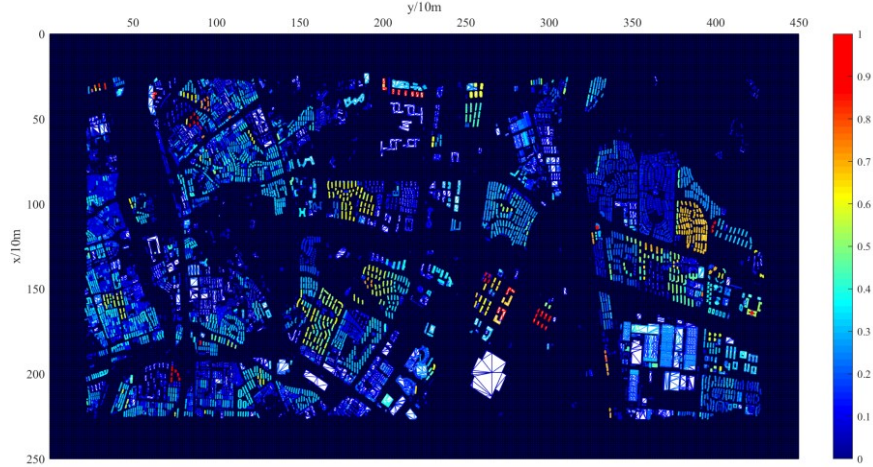

(b) h=160m

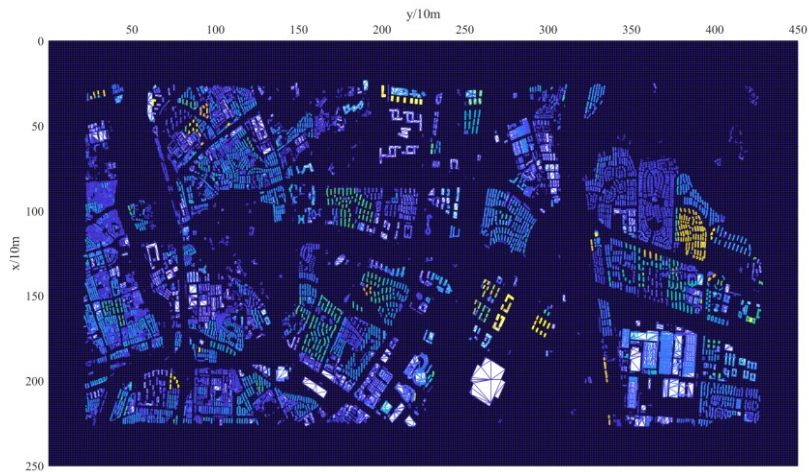

(c) h=220m

Figure 20 Urban airspace environment after risk assessment with different $h$

The adjustment of the weight coefficients can reflect actual conditions or the preferences of managers. In densely populated areas, along highway segments, or at UAV delivery points, the corresponding weights $\omega_1$, $\omega_2$ and $\omega_3$ for ground personnel, ground vehicles, and UAVs, respectively, can be increased as needed.

**6.3.2 Impact of cell size on the single-aircraft track planning**

In the predefined urban low-altitude airspace environment and existing tracks, the effect of varying cell division sizes on track planning outcomes was analyzed. The results, summarized in Table 4, indicate that a finer cell size reduces operational risk but increases the number of waypoints, suggesting more sophisticated track planning. Moreover, transportation cost and computational time increase. This suggests that a smaller cell size enhances the simulation of buildings and risk assessment in the urban airspace environment, offering a more detailed representation. Such granularity may unlock additional airspace and provide aircraft with more flexibility in course adjustments, potentially reducing track length. Nonetheless, this comes at the expense of increased computational load, indicating that while a finer cell size offers a safer and potentially more efficient route, it also requires more resources in terms of computational time. Furthermore, the physical limitations in the turning capability of the aircraft might result in significant deviations between the actual and planned track. In contrast, a larger cell size provides a coarser representation of buildings and risk assessment but improves computational efficiency. Therefore, selecting an appropriate cell size based on actual airspace conditions and aircraft performance is crucial for balancing operational safety, efficiency, and computational requirements.

Table 4 Comparisons of optimal tracks with different cell sizes

| Index | 10m×10m×10m | 5m×5m×5m | Rate of the changes |
|---|---|---|---|
| Operational risk | 196 | 163.33 | -16.67% |
| Transportation cost(￥) | 27.1527 | 25.7975 | -5.05% |
| Number of waypoints | 386 | 636 | 64.89% |
| Computational time (s) | 68.71 | 659.52 | 859.22% |

The calibration of the weight coefficients $\omega_4$ for operational risk cost and $\omega_5$ for transportation cost is important to reconciling operational efficiency with cost-effectiveness in track planning within a complex urban environment. Thus, determining the right balance between these two competing factors is essential for achieving the desired trade-off between safety and efficiency in track planning.

As shown in Figure 21, the Pareto curve demonstrates the trade-offs between operational risk and transportation cost. Each point on the curve represents an optimal solution derived from different weight combinations, where reducing operational risk leads to an increase in transportation cost, indicating their interdependence. Decision-makers can find the most suitable balance between risk control and cost efficiency based on their needs.

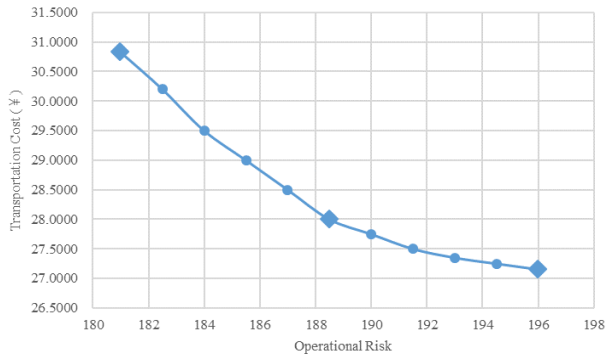

Figure 21 Trade-off analysis between operational risk and transportation cost

### 6.3.3 Impact of operational parameters on the multi-aircraft scheduling optimization

With a fixed number of aircraft and a constant total number of planned flights, the outcomes of scheduling optimization under various sets of flight numbers and speeds were analyzed. The data indicates that optimization reduces delay times, irrespective of the number of planned flights.

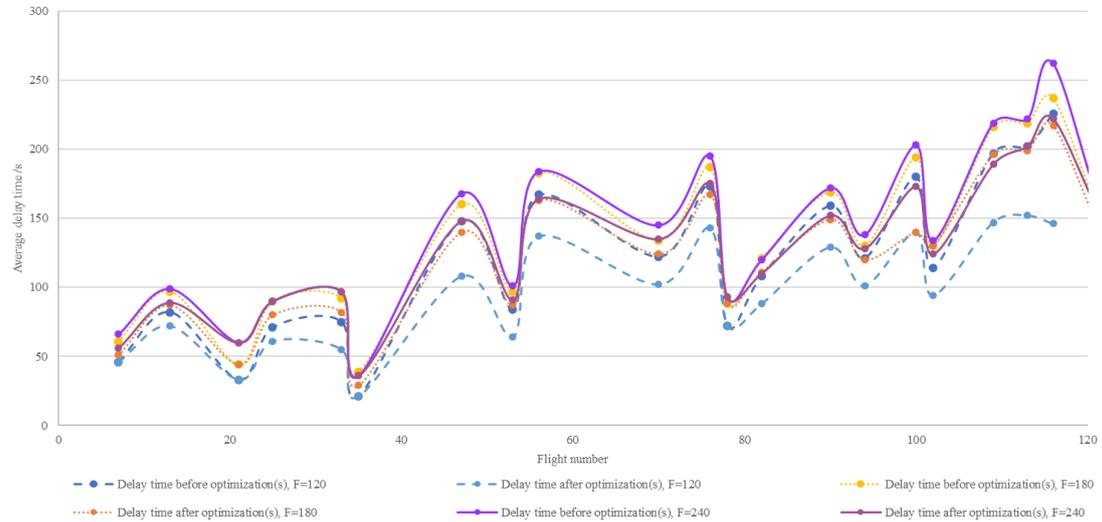

Figure 22 Average delay time before and after optimization with different $F$

As displayed in Figure 22, when the total number of planned flights is low, the airspace operates at a moderate level of activity, resulting in higher task fulfillment rates and shorter average delays per flight. Conversely, as the total number of planned flights increases, the airspace becomes more congested, leading to lower task fulfillment rates and extended average delays per flight. Strategic scheduling optimization can effectively reduce delay times, but its effectiveness is influenced by the number of flights in the system. Fewer planned flights yield more efficient operations, while an increase in planned flights presents challenges that may affect the overall efficiency and timeliness of the airspace system.

The impact of varying speeds on average delay times is illustrated in Figure 23. Slower speeds correlate with longer cell occupation times, restricting maneuverability and increasing average delay times. Conversely, higher speeds reduce the time spent in each cell, improving maneuverability and decreasing flight delays. This suggests that higher speeds enhance the effectiveness of the optimization process, significantly reducing delays and improving the overall efficiency of the air traffic management system.

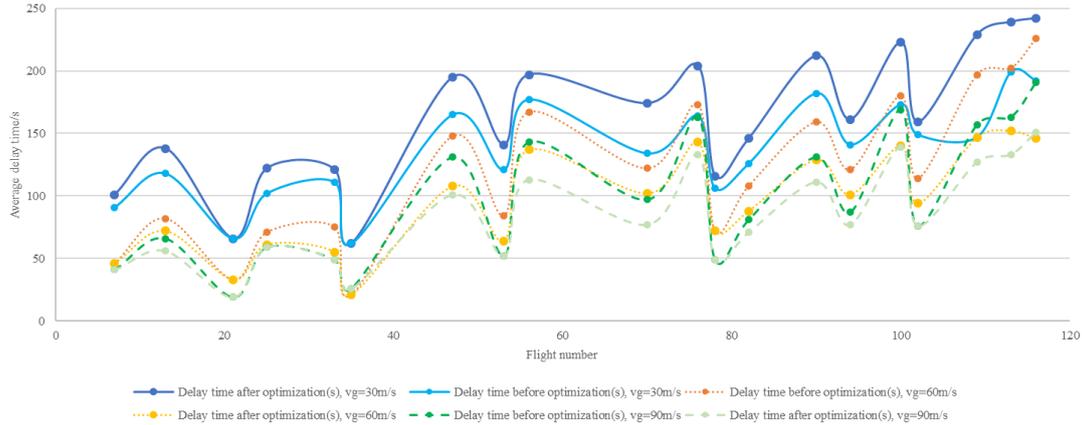

Figure 23 Average delay time before and after optimization with different $v_g$

## 7. Conclusions

This research aims to improve the operational safety and efficiency of UAM systems by developing a trajectory-based scheduling optimization framework. Initially, a multi-layer urban airspace environment model is constructed, capturing potential aircraft failure scenarios and their impacts on ground personnel, vehicles, and small multi-rotor UAVs. Subsequently, a single-aircraft track planning approach is formulated, balancing operational risk and transportation costs for eVTOL aircraft. An algorithm including safety buffer zones is designed to generate solutions that prevent potential safety clearance breaches at future waypoints along the estimated track. Lastly, a multi-aircraft scheduling optimization model is devised to minimize average delay time and ensure the avoidance of dynamic air collisions. A scheduling decision process-driven algorithm is integrated into the solution to mitigate airspace congestion. Simulation experiments validate the effectiveness of the proposed model and algorithms, while sensitivity analysis demonstrates that flight altitude, cell size, and operational parameters such as the total number of flights and aircraft speed influence the outcomes to varying degrees. Compared to previous studies [48-50], this evaluation index system offers a more thorough approach, accounting for static risks (such as terrain variation and building distribution), dynamic airborne collision risks (involving small multi-rotor UAVs), and ground collision risks (involving personnel and vehicles).

This work contributes methodologies that could enhance the safety and efficiency of UAM operations, aligning with the broader vision of intelligent transportation as outlined in the AAM concept. As UAM technology evolves, assumptions such as the uniform distribution of ground personnel and vehicles, as well as constant speeds for small multi-rotor UAVs, could be gradually relaxed. In practice, fluctuating traffic densities and varying flight speeds are expected and can be accounted for in future models. Moreover, the current track planning approach addresses only the cruise phase between origins and destinations. Future research could include vertical take-off and landing paths, especially in dense urban environments. Simultaneously, while the present model is based on regular environmental conditions, future work could integrate real-time meteorological data, such as wind velocity, precipitation, or snowfall, to adjust flight paths dynamically. Furthermore, by implementing advanced conflict resolution strategies—such as velocity adjustments or holding patterns—future studies can handle more complex, real-world scenarios, optimizing airspace usage and improving overall transportation performance.

**CRediT authorship contribution statement**

**Jin Zhang**: Data curation, Investigation, Methodology, Software, Validation, Writing – original draft. **Xiaoran Qin**: Conceptualization, Funding acquisition, Methodology, Resources, Supervision, Writing - review & editing. **Ming Zhang**: Conceptualization, Funding acquisition, Project administration, Resources, Supervision.

**Declaration of Competing Interest**

The authors declare that they have no known competing financial interests or personal relationships that could have appeared to influence the work reported in this paper.

**Acknowledgments**

The work described in this paper is partially supported by the General Program of the National Natural Science Foundation of China (No.52272350) and National Natural Science Foundation of China (No.72301111).

## Appendix A

Following the principles of dynamics, the impact kinetic energy $E_{imp}$ of the falling aircraft is known as:

$$E_{imp} = \frac{1}{2}(m_e + m_p)v^2 \quad \text{Eq. (A.1)}$$

where the values of $m_e$ and $m_p$ are given, representing the mass of the aircraft and the passengers respectively. $v$ is the velocity at the moment of ground contact.

Due to the influence of gravity and air resistance on the falling process of the aircraft, it exhibits a variable accelerated motion, where the velocity $v$ is the integral of the acceleration $a$, i.e., $v = \int_0^t a\,dt$. The corresponding calculation of $a$ takes the form as:

$$a = \frac{F_g - F_z}{m_e + m_p} = \frac{F_g - \frac{1}{2}R_I\rho_a S_e v_z^2}{m_e + m_p} = \frac{2(m_e + m_p)g - R_I\rho_a S_e v_z^2}{2(m_e + m_p)} = g - \frac{R_I\rho_a S_e v_z^2}{2(m_e + m_p)} \quad \text{Eq. (A.2)}$$

Among the forces acting on the aircraft, $F_g$ corresponds to the gravitational force. Likewise, $F_z$ specifies the formula for air resistance in physics, where $R_I$ is the air resistance coefficient, $\rho_a$ is the air density, $S_e$ is the frontal area of the aircraft (equal to the area of the falling zone), and $v_z$ is the relative velocity between the aircraft and the air.

By substituting the expression of $a$ into $v$, the following is obtained:

$$v = \int_0^t \left(g - \frac{R_I\rho_a S_e v_z^2}{2(m_e + m_p)}\right)dt = \sqrt{\frac{2(m_e + m_p)g}{R_I\rho_a S_e}\left(1 - e^{-\frac{h'R_I\rho_a S_e}{m_e + m_p}}\right)} \quad \text{Eq. (A.3)}$$

where $h'$ indicates the altitude of the aircraft above ground personnel.

As a result, the risk posed to small multi-rotor UAVs due to the aircraft fall can be articulated as:

$$R^U = \lambda \rho_U V_e \quad \text{Eq. (A.4)}$$

Assuming the aircraft flies along the $x$-axis and considering its variable speed motion in that direction, the volume of space swept by the collision box $V_e$ can be stated as:

$$V_e = e_w e_h (v_r' t + e_l) \quad \text{Eq. (A.5)}$$

where $v_r'$ designates the average relative velocity between the aircraft and the UAV. Based on the findings from Gao et al. [51], assuming the velocity direction of the UAV is uniformly distributed, $v_r'$ is calculated as an average quantity across the ranges of time $t$, aircraft velocity $v_e$, and angles $\gamma$ and $\varepsilon$, obtained by a quadruple integral over $v_r$. It can be formulated as:

$$v_r' = \frac{1}{\pi^2 t(v_{e\,max} - v_{e\,min})} \int_0^t \int_{v_{e\,min}}^{v_{e\,max}} \int_{-\frac{\pi}{2}}^{\frac{\pi}{2}} \int_0^{\pi} v_r \, d\varepsilon \, d\gamma \, dv_e \, dt \quad \text{Eq. (A.6)}$$

where $v_r$ is used for the relative velocity between the aircraft and the UAV, being decomposed into horizontal relative velocity $v_{rx'y'}$ and vertical relative velocity $v_{rz'}$ for separate calculations, known as:

$$v_{rx'y'} = \sqrt{(v_e \cos\theta)^2 + (v_U \cos\gamma)^2 + 2v_e \cos\theta\, v_U \cos\gamma \cos\varepsilon} \quad \text{Eq. (A.7)}$$

$$v_{rz'} = v_e \sin\theta - v_U \sin\gamma \quad \text{Eq. (A.8)}$$

where $v_e$ and $v_U$ are the velocities of the aircraft and UAV, $\theta \epsilon \left(-\frac{\pi}{2}, \frac{\pi}{2}\right)$ and $\gamma \epsilon \left(-\frac{\pi}{2}, \frac{\pi}{2}\right)$ the angles relative to the $xy$ plane, and $\varepsilon \epsilon (0, \pi)$ the angle between $v_e$ and $v_U$. The expression of $v_r$ is as follows:

$$v_r = \sqrt{v_{rx'y'}^2 + v_{rz'}^2} = \sqrt{v^2_e + v^2_U + 2v_e v_U (\cos\theta \cos\gamma \cos\varepsilon - \sin\theta \sin\gamma)} \qquad \text{Eq. (A.9)}$$

**Appendix B**

Table B.1 Values of shielding coefficient $c_s$[1]

| Geomorphological condition | $c_s$ |
|---|---|
| No obstacles | 0 |
| Sparse trees | 0.25 |
| Trees and low buildings | 0.50 |
| High buildings | 0.75 |
| Industrial buildings | 1 |

Table B.2 UAS Operations Risk Reference System [52]

|  |  | Catastrophic [a] | Hazardous [b] | Major [c] | Minor [d] | No safety effect |
|---|---|---|---|---|---|---|
| Frequent | >10⁻³ | ▓ | ▓ | ▓ |  |  |
| Probable | <10⁻³ | ▓ | ▓ | ▓ |  |  |
| Remote | <10⁻⁴ | ▓ | ▓ |  |  |  |
| Extremely remote | <10⁻⁵ | ▓ |  |  |  |  |
| Extremely Improbable | <10⁻⁶ |  |  |  |  |  |

[a] Uncontrolled flight and/or uncontrolled crash, which can potentially result in a fatality. Potential fatality to UAS crew or ground staff.

[b] Controlled-trajectory termination or forced landing potentially leading to the loss of the UAS where it can be reasonably expected that a fatality will not occur. Potential serious injury to UAS crew or ground staff.

[c] Emergency landing of the UAS on a predefined site where it can be reasonably expected that a serious injury will not occur. Potential injury to UAS crew or ground staff.

[d] Slight reduction in safety margins or functional capabilities and slight increase in UAS crew workload.

Table B.3 Parameters of simulation experiment[2]

| Parameter | Value | Parameter | Value |
|---|---|---|---|
| $\lambda$ | 6.04×10⁻⁵/h | $h_{min}$ | 90m |
| $\rho_P$ | 2.5×10⁻⁴/m² | $h'_{min}$ | 30m |
| $d$ | 6m | $h_{max}$ | 3000m |
| $c_s$ | $\begin{cases} 0.5, \text{ height of buildings} \leq 15m \\ 0.75, \text{height of buildings} >15m \end{cases}$ | $h'_{max}$ | 300m |
| $\alpha$ | 10⁶J | $s_{min}$ | 50m |
| $\beta$ | 232J | $L_e$ | 30km |
| $m_e$ | 400kg | $\omega_4$ | 0.5 |
| $m_{p\,max}$ | 220kg | $\omega_5$ | 0.5 |
| $g$ | 9.8m/s² | $f$ | 120 |
| $R_I$ | 0.3 | $T_D$ | 240s |

---

[1] Hu, X., Pang, B., Dai, F., and Low, K. H., 2020. Risk assessment model for UAV cost-effective path planning in urban environments. IEEE Access, 8, 150162-150173.

[2] Performance parameters of the aircraft are from the National Aviation Database: https://www.aviationdb.com. Accessed on April 23, 2024.

| | | | |
|---|---|---|---|
| $\rho_a$ | 1.225kg/m³ | $T_R$ | 360s |
| $h'$ | 148.2m | $\chi$ | 0.3 |
| $S_v$ | 9.68m² | $L_g$ | 4365m |
| $\rho_V$ | 0.07/m | $v_g$ | 25m/s |
| $w_r$ | 3.5m | $\Delta T_a$ | 10s |
| $\rho_U$ | 3.48×10⁻⁸/m³ | $T_0$ | 100°C |
| $e_l$ | 5.63m | $\xi$ | 0.99°C/min |
| $e_w$ | 5.63m | $T_F$ | 0.1°C |
| $e_h$ | 1.855m | $l$ | 200 |
| $v_{e\,min}$ | 10m/s | $V_1$ | (30,30,0) |
| $v_{e\,max}$ | 130km/h | $V_2$ | (30,422,0) |
| $v_U$ | 6m/s | $V_3$ | (222,30,0) |
| $\theta$ | $\dfrac{\pi}{6}$ | $V_4$ | (222,422,0) |
| $\gamma$ | $\dfrac{\pi}{2}$ | $V_5$ | (126,226,0) |
| $\varepsilon$ | $\dfrac{\pi}{2}$ | $L_{12}$ | 3920m |
| $\omega_1$ | 0.5 | $L_{13}$ | 1920m |
| $\omega_2$ | 0.3 | $L_{14}$ | 4365m |
| $\omega_3$ | 0.2 | $L_{15}$ | 2183m |
| $P_O$ | (30,30,120) | $L_{23}$ | 4365m |
| $P_D$ | (222,422,120) | $L_{24}$ | 1920m |
| $c_h$ | 5.135×10³ J/m | $L_{25}$ | 2183m |
| $c_v$ | 4.65×10⁵ J/m | $L_{34}$ | 3920m |
| $c_e$ | ¥5.96×10⁻⁷/J | $L_{35}$ | 2183m |
| $\tau_{max}$ | 1.56 | $L_{45}$ | 2183m |
| $m_{max}$ | 650kg | $\omega_6$ | 0.6 |
| $v_{we}$ | 26.45m/s | $\omega_7$ | 0.4 |
| $\mu$ | 100 | | |